\documentclass[a4]{article}
\usepackage{natbib}
\usepackage{graphicx}
\usepackage{color} 

\title{Epigenetic Tracking,\\
       a Method to Generate Arbitrary Shapes\\ 
       By Using Evolutionary-Developmental Techniques}
       
\author{Alessandro Fontana\\
   \rm  IEEE\\
   \rm  alessandro.fontana@ieee.org
   }
            
\begin{document}
\maketitle

\begin{abstract}
%\colorbox{red}{intro}
This paper describes an Artificial Embryology method (called ``Epigenetic Tracking'') to generate predefined arbitrarily shaped 2-dimensional arrays of cells by means of evolutionary techniques. It is based on a model of development, whose key features are: i) the distinction bewteen ``normal'' and ``driver'' cells, being the latter able to receive guidance from the genome, ii) the implementation of the proliferation/apoptosis events in such a way that many cells are created/deleted at once, in order to speed-up the morphogenetic process. iii) the presence in driver cells of an epigenetic memory, that holds the position of the cell in the cell lineage tree and represents the source of differentiation during development. The experiments performed with a number of 100x100 black and white and colour target shapes (the horse, the couple, the hand, the dolphin, the map of Britain, the foot, the frog, the baby, the stomach, the french flag, the head) bring to the conclusion that the method described is able to generate any target shape, outperforming any other known method in terms of size and variety of the generated shapes. The interpretation of the proposed method as a model of embryogenesis and its biological implications are discussed.
\end{abstract}

\section{Introduction and Related Work}

%\colorbox{red}{intro}
This paper belongs to the field of Artificial Embryology and more specifically addresses the problem of morphogenesis. It describes a method, called ``Epigenetic Tracking'' (``Cell Tracking'' in previous work \citep{FX07}), to generate arbitrarily shaped 2-dimensional arrays of cells by means of evolutionary-developmental techniques, i.e. by evolving genomes that guide the development of the shape starting from a single cell. The paper is organised as follows: the rest of this section surveys the related work, section 2 describes the model of development, section 3 describes the experiments performed, section 4 discusses the biological implications, section 5 draws the conclusions and outlines future work.

%\colorbox{yellow}{AE related work}
The previous work in the field of Artificial Embryology (see \citep{KB03,SM03} for a comprehensive review) can be divided into two broad categories: the grammatical approach and the cell chemistry approach. The grammatical approach, originated by Lindenmayer \citep{LX68}, evolves sets of rules in the form of grammatical rewrite systems; the grammar can be context-free or context-sensitive and can utilise parameters; variations on this theme include using instruction trees or directed graphs in place of actual grammars. L-systems were employed as a means of describing the complex fractal patterns observed in nature and particularly the architecture of plants. The cell chemistry approach draws inspiration from the early work of Turing \citep{TX52}, who introduced a mathematical model of diffusion and reaction within a physical substrate. This approach attempts to mimic more closely how physical structures emerge in biology; cells are arranged in a physical space where simulated proteins can be sent as signals from one cell to another, as in nature.

%\colorbox{yellow}{AE grammatical approach}
Within the grammatical approach, Sims \citep{SX94} used directed graphs to evolve the body morphologies and neural networks of artificial creatures in a simulated 3D physical world; in these graphs, a node represents a body part and an edge specifies how body parts are connected. Using a domain similar to Sims', Hornby and Pollack \citep{HP02} applied L-systems to the simultaneous evolution of the body morphologies and neural networks of artificial creatures in a simulated 3D physical environment. Cangelosi, Nolfi and Parisi \citep{CN94} devised a model of neural development which includes cell division and cell migration in addition to axonal growth and branching; the development process shows successive phases of functional differentiation and specialisation. Gruau's Cellular Encoding \citep{GW96} uses grammar trees to encode steps in the development of a neural network starting from a single ancestor cell; the grammar tree contains developmental instructions at each node.

%\colorbox{yellow}{AE-cell chemistry approach}
Within the cell chemistry approach, Random Boolean Networks (RBN's) were originally developed by Kaufmann as a model of genetic regulatory networks \citep{KF69}; in the context of the development of multi-cellular organisms, the attractors of RBN's are interpreted as the different ``cell types'' of the organism. Dellaert and Beer \citep{DB96} presented models of development to evolve functional autonomous agents, complete with bodies and neural control systems. De Garis \citep{DG99} developed a model for evolving shapes in 2D reproductive cellular automata; the model was successful in evolving convex shapes but non-convex shapes (e.g. the L-shape) presented a problem. Bongard and Pfeifer \citep{BP01} proposed a minimal model of ontogenetic development to evolve both the morphology and neural control of agents that perform a block-pushing task in a physically-realistic, virtual environment. Inspired by the cell adhesion process, Hogeweg \citep{HX03} developed a model to simulate morphogenetic processes such as cell migration or engulfing, achieving to evolve complex artificial organisms. Miller and Banzhaf \citep{MB03} developed artificial organisms (the french flag) based on a method called Cartesian Genetic Programming, which evolves a developmental program inside cells. Geard and Wiles \citep{GW03} used a simple recurrent network to model the process of gene regulation and evolved systems that were able to generate the first four cell divisions of the C. elegans cell lineage tree. Eggenberger \citep{EX97,EX04} used artificial cells endowed with genetic regulatory networks to evolve and develop simulated creatures; by using developmental mechanisms such as asymmetric cell division, genetic regulation and cell adhesion and physical interactions between cells, he achieved to shape multicellular (moving) 3d organisms. Roggen, Federici and Floreano \citep{FF07} investigated the potential of a very simple intrinsic, online and cellular developmental system designed for multi-cellular circuits called morphogenetic system, inspired by gene expression and cellular differentiation and focused on low computational requirements and a compact hardware implementation.

%\colorbox{red}{question to address}
The task of generating predefined arbitrary shapes has so far proved to be a difficult one. The evolved shapes reported in the literature are often very simple and of small size (e.g. the french flag); moreover, some of the algorithms used to generate them seem to contain ad-hoc solutions that bias them towards certain shapes or categories thereof. The method described in this paper tries to overcome these limitations and provide a general solution to the problem of generating arbitrary shapes of arbitrary size.  

\section{The Model of Development}

%\colorbox{yellow}{space, time, development, cell variables}
In our model of development the phenotype of the organism is represented as a 2-dimensional array of square-shaped cells, being each cell associated to a position on a grid. The development starts with a single cell placed in the middle of the grid, and unfolds in n development phases, counted by the variable ``global development phase'' (GDP) that runs from 0 to n-1 (n is a parameter). The term ``global'' refers to the fact that the variable GDP is shared by all cells (and therefore it can be considered the global ``clock'' of the organism). To each cell four variables are associated:

\begin{itemize}
\item a flag indicating whether the cell is ``driver'' or ``normal''; 
\item the ``genome'', organised as an array of ``development operators'', which is identical in all cells;
\item the ``cellular epigenetic type'' (CET), organised as an array of n integers (n is the number of development phases), which is not identical in all cells; the CET is present only in driver cells; 
\item an integer representing the cell's colour. In the current implementation four colour values are foreseen (0,1,2,3), an extra value (-1) indicates cell absence; 
\end{itemize}

%\colorbox{yellow}{driver and normal cells, CET, proliferation basics}
Cells belong to two categories: ``driver'' cells (coloured in yellow in the figures) and ``normal'' cells (coloured in orange or blue). The basic difference between driver and normal cells is that the first can be instructed by the genome (by means of an operator whose left part matches the CET value of the cell) to proliferate or induce apoptosis in the surrounding area. Figure 1 shows an example: a driver cell associated to a CET value labelled with ``A''(called ``mother cell'') proliferates in an area around it (called ``development area'', delimited by a dotted line in the figure). While proliferating, it mostly generates normal cells (which fill the development area) and other driver cells, which are much fewer in number and ``dot'' the development area).

\begin{figure}[ht]
\begin{center}
\includegraphics[width=.60\columnwidth]{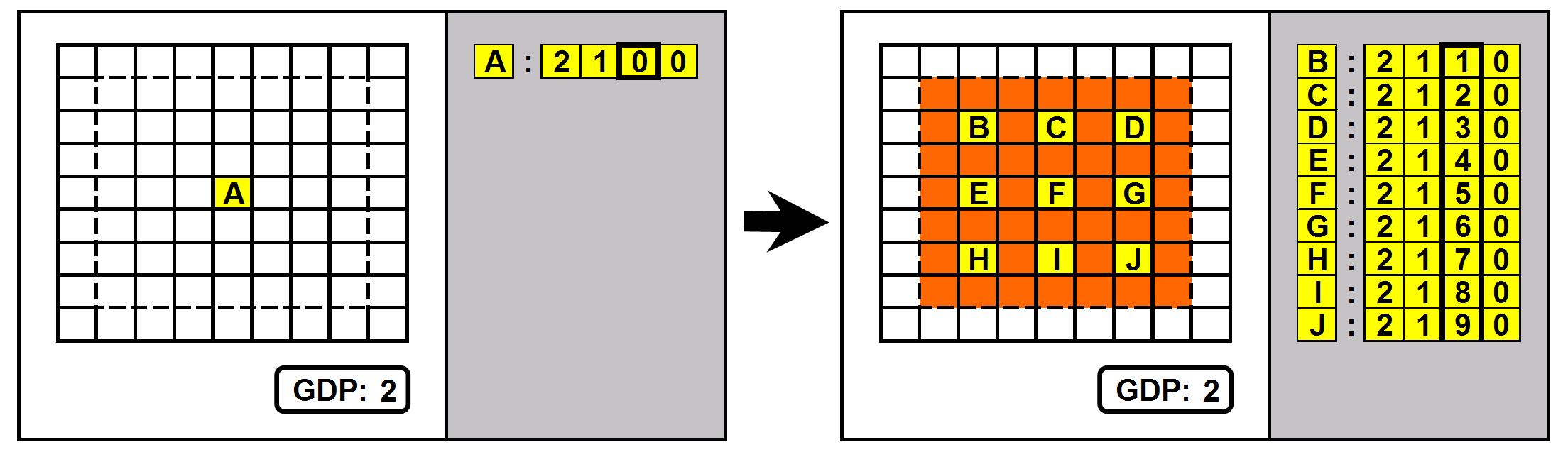}
\caption{Proliferation event. The genome triggers a driver cell to proliferate in an area around it, generating normal and driver cells.}
\label{00Basic}
\end{center}
\end{figure}

%\colorbox{red}{assignment of CET values}
A key point is the mechanism of assignment of the CET values on the newly created driver cells. To each new driver cell a new CET value is assigned, starting from the mother cell's CET value (the array [2100] in the figure, labelled with ``A'') and adding 1 to the value of the i-th position of the array at each new assignment, where i is the current GDP value (2 in the figure, the column corresponding to the GDP value is highlighted with a thicker border); with reference to the figure, the new driver cells are assigned the values [2110],[2120],... , labelled with ``B'',``C'', etc. In practise the variable CET holds the position of the driver cell in the driver cell lineage tree: this ensures that the new CET values are all different from the mother's value and from each other. Whether one of these new CET values will become the centre of another proliferation event depends on the presence in the genome of an operator whose left part matches such value. 

\begin{figure}[ht]
\begin{center}
\includegraphics[width=.60\columnwidth]{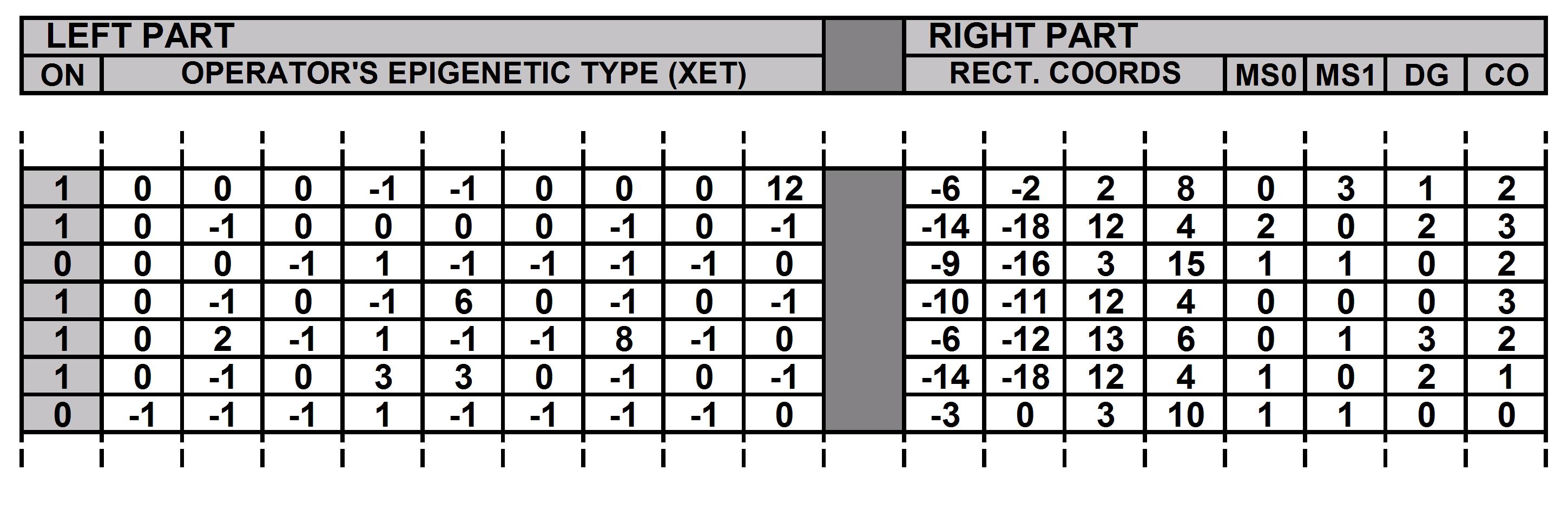}
\caption{genome, development operators.}
\label{00Genome}
\end{center}
\end{figure}

%\colorbox{yellow}{genome}
The genome as we said is organised as an array of development operators (see figure 2). Excluding mutations (not dealth with in this work), the genome is not modified during development and is identical in all cells. Each development operator has a left part and a right part. The left part consists of a variable called XET, having the same structure of the variables CET: if the XET value is equal to the CET value of a given driver cell, the operator is activated and the relevant code specified in the right part is executed for that cell; An additional flag indicates whether the operator is ``structurally'' inactive or not. The right part of the operator has:

\begin{itemize}
\item a field with the coordinates of the rectangle which delimits the development area (row and column values of the north-west and south-east corners of the rectangle)
\item a field holding a ``master switch'' (MS0) that defines the shape of the development area (``rectangular'' -value = 0, ``diagonal left'' -value = 1, ``diagonal right'' -value = 2)
\item a field holding a second ``master switch'' (MS1) that defines the type of ``development event'' that is going to occur (``proliferation'' -value=0, ``apoptosis'' -value=1).
\item a field with a parameter that specifies the ``thickness'' of the diagonal (valid only if MS0=1 or 2)
\item a field with a parameter that specifies the colour of the newly created cells (both normal and driver)
\end{itemize}

\begin{figure}[ht]
\begin{center}
\includegraphics[width=.60\columnwidth]{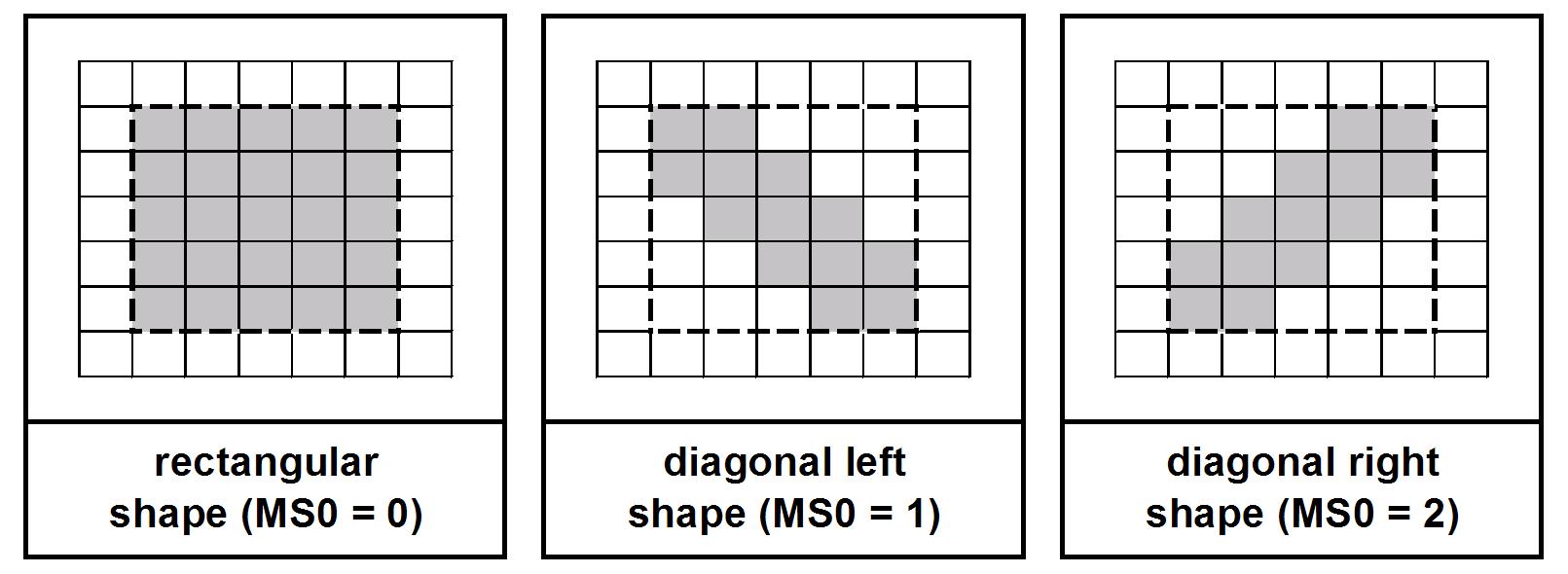}
\caption{Effect of the master switch MS0 on the development area}
\label{figx}
\end{center}
\end{figure}

%\colorbox{yellow}{proliferation and apoptosis}
In case of proliferation the development area is filled with newly created cells: most of the cells generated are normal cells, some are driver cells. The driver cells are much fewer in number (usually a ``linear normal to driver ratio'' of 5 has been used, corresponding to a 2-dimensional ratio of 5*5=25) and are deployed evenly on the development area: they appear as yellow dots in the figures. In case of apoptosis, the mother cell and all the cells contained in the development area ``die'', i.e. are deleted from the grid. The different types of development events correspond to the tools of a painter: the proliferation corresponds to the paintbrush and the apoptosis corresponds to the eraser.

\begin{figure}[h!]
\begin{center}
\includegraphics[width=.60\columnwidth]{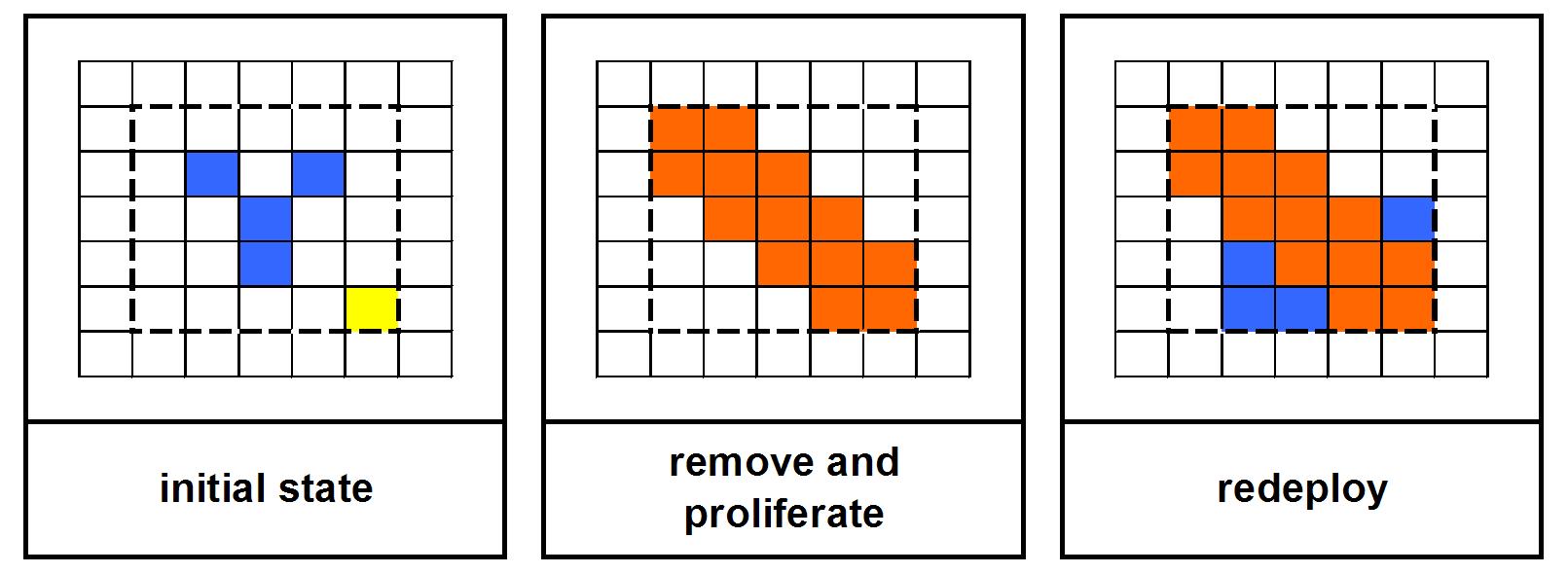}
\caption{Remove-redeploy mechanism, shown for the NW quadrant (newly generated cells in orange, old cells in blue).}
\label{00RemoveRedeploy}
\end{center}
\end{figure}

%\colorbox{yellow}{remove-redeploy mechanism}
Special attention is required if the development area is not empty. In this case the cells present must be either moved to other locations in the grid or removed altogether (overwritten). The solution chosen consists of first removing the cells present in the development area, carrying out the proliferation and finally redeploying the cells removed to the first empty positions available, starting from the position of the mother cell and going outwards. This procedure is carried out for each quadrant of the development area (first NW, then NE, SE and finally SW).

%\colorbox{yellow}{''external'' driver cells}
After the application of the remove-redeploy mechanism, it may happen that the borders of the developing shape remain devoid of driver cells, which would make subsequent development much harder. The solution adopted is to add another ''sprinkle'' of driver cells on shape borders at the end of each development step, after the application of all development operators. The generation of such ''external'' driver cells is attributed to the nearest ''internal'' driver cell (due to computational constraints the Manhattan distance is used). 

%\colorbox{yellow}{ET counter}
Due to computational constraints, the CET assignment mechanism has been simplified through the use of a scalar CET and the introduction of a global CET counter. When a new driver cell is created by any proliferation event, the CET value assigned corresponds to the value taken by the CET counter at the beginning of the proliferation event. The CET counter has the value 1 at the beginning of development and is incremented by one at each new assignment (0 is the CET value of the zygote): in practise the CET value is a progressive number representing the creation order.

\begin{figure}[t]
\begin{center}
\includegraphics[width=.60\columnwidth]{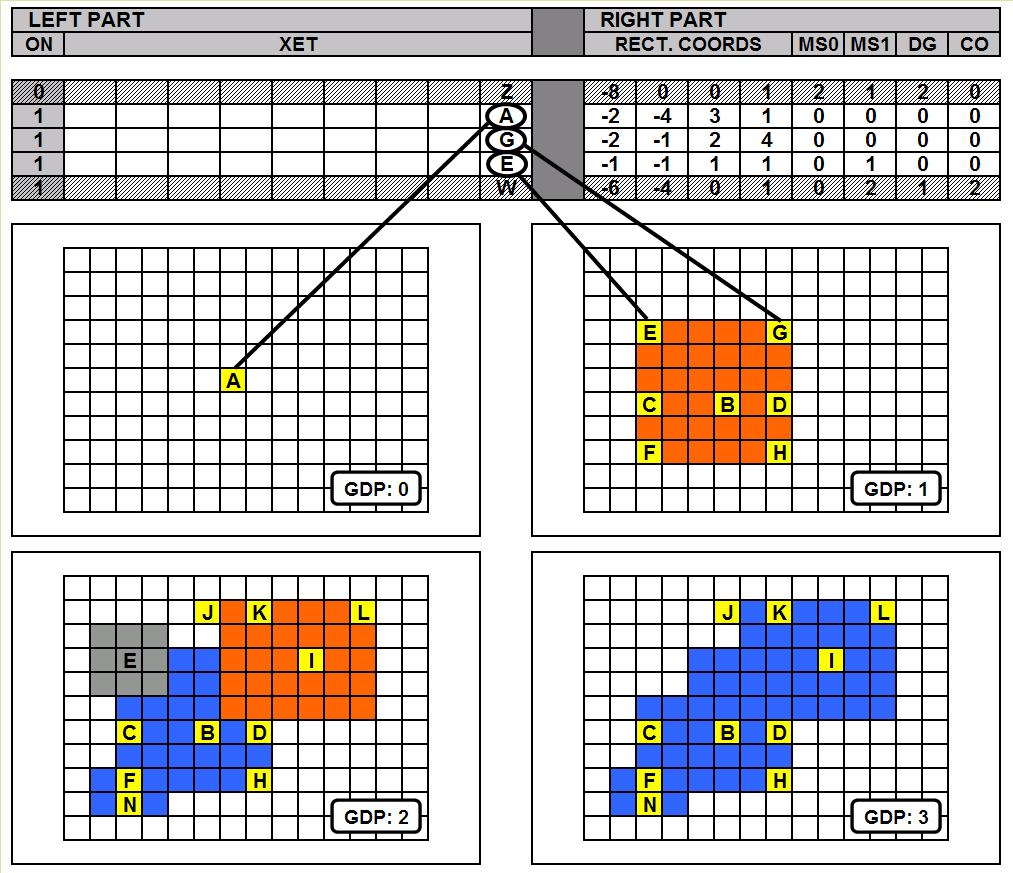}
\caption{Example of development in four phases, driven by three operators (normal newly generated cells in orange, normal old cells in blue, driver cells in yellow). Positioning of driver cells is performed by the algorithm used for the experiments, while in figure 1 it is done manually, as an example.}
\label{00Devex}
\end{center}
\end{figure}

%\colorbox{yellow}{example of development phases}
Figure \ref{00Devex} shows an example of development in four development phases (GDP=0,1,2,3) steered by three development operators, the first (a rectangular proliferation) triggered by the CET value labelled ``A'' in phase 0, the second (a rectangular apoptosis) triggered by the CET value labelled ``E'' in phase 2, the third (a rectangular proliferation) triggered by the CET value labelled ``G'' also in phase 2. The CET value ``A'' was present at the beginning of development, The CET values ``E'' and ``G'' have been created in phase 1.   

%\colorbox{red}{key feature 1}
Before presenting the outcomes of the experiments performed, we wish to highlight the key features of the model of development described. The first key feature of our model is the distinction between normal cells and driver cells. Only driver cells have a CET value and can therefore be instructed to develop (proliferate or die) by the genome; they represent the scaffolding, the backbone of the developing shape and make possible to steer the development of the whole shape by acting on a small subset of cells. If all cells (both driver and normal) had an associated CET value, the space the GA would have to search would be unmanageable.

%\colorbox{red}{key feature 2}
The second feature is the implementation of the development events of proliferation and apoptosis in such a way that they create/delete many cells at once (instead of one). This increases the power of the single development event, allows a reduction of the number of development operators needed to generate a given shape and has the ultimate effect of speeding-up considerably the morphogenetic process. Together with the previous one, this feature serves the purpose of reducing the number of cells the genome has to steer, at the same time endowing such cells with a capacity to influence more profoundly the course of development.

%\colorbox{red}{key feature 3}
The third feature of the model is the explicit presence of an epigenetic memory, i.e. a cell variable (the CET, only present in driver cells) that takes different values in different cells and represents the source of differentiation during development, leading different cells at different times to read out and execute different parts of the genome. It is by means of the cell epigenetic type that driver cells know what type of cells they are and how their behaviour has to be.

%\colorbox{red}{key feature 4}
The fourth feature is the mechanism of assignment of the CET values on the newly generated driver cells during a proliferation event. Just as a mother gives each of her newborn babies a distinct ``name'', such mechanism ensures that each new driver cell is assigned a new, previously unseen CET value; the CET value, corresponding to the position of the driver cell in the driver cell lineage tree, represents the link by which these driver cells in subsequent phases can be picked up by the genome and given other instructions to execute. If new driver cells were not guaranteed to have a distinct name, the genome would not be able to pick them individually: as a consequence, the GA would not be able to explore freely all regions of the space and the developmental trajectories would therefore be biased towards certain regions of the space, making the development of arbitrary shapes hard.

%\colorbox{red}{key feature 5}
The fifth feature, that descends from the latter two features, is the ``looseness'' of the driver cell lineage tree.   
It is important to point out that, when new CET values are generated in the course of a proliferation event, the mother cell does not ``know'' in advance if these CET values will activate a development operator in the genome: such an operator does not need to be there. Lacking this information, the most reasonable thing to do is to assign the daughter cells distinct names, creating the potential for a match to occur. The byproduct of this approach is that at any given moment in the course of evolution there will be many driver cells that do not activate any operators during development and, at the same time, many operators that are not activated by any driver cell: the presence of inactive driver cells and inactive operators is an unavoidable characteristic of the model proposed (figure 5 illustrates this concept): we believe that such ``looseness'' increases the model's evolvability. We claim that this combination of factors is unique to our model and allows to reach a high level of performance in terms of both size and variety of the evolved shapes.

\begin{figure}[!t]
\begin{center}
\includegraphics[width=0.60\columnwidth]{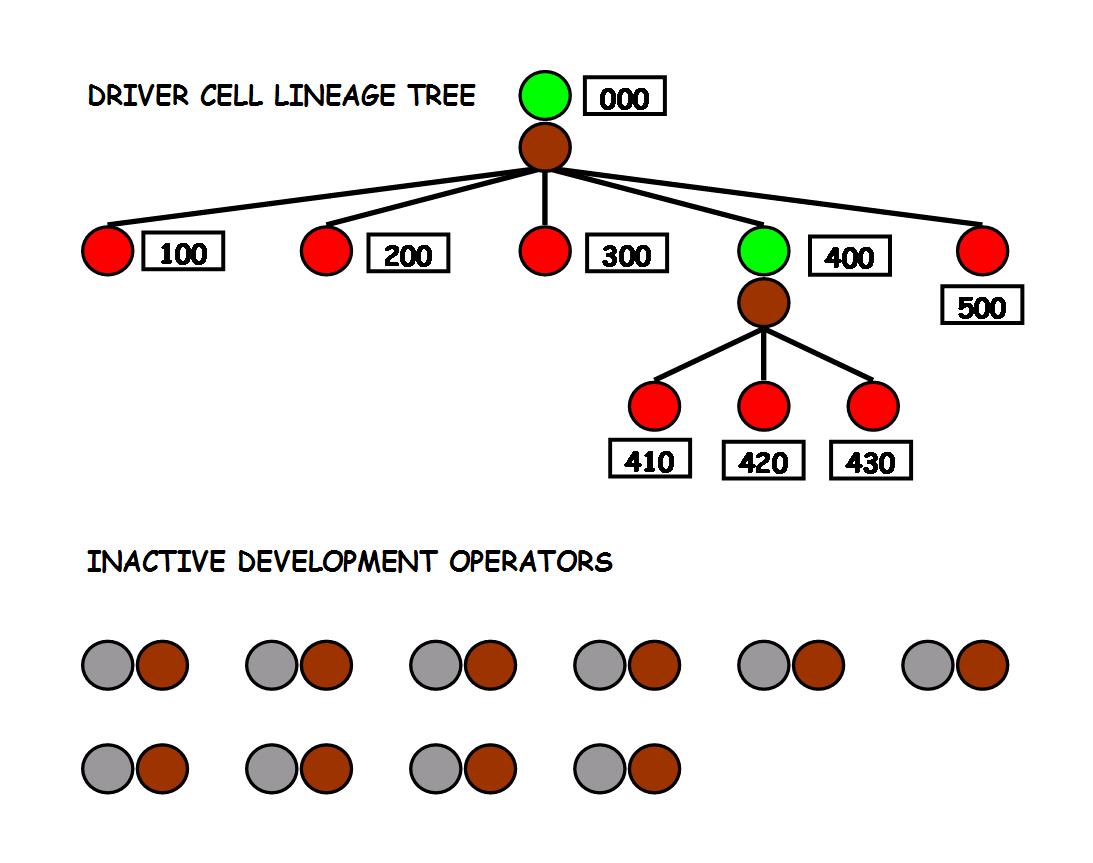}
\caption{The driver cell lineage tree. In the upper part the lineage tree of drivers cells; red circles represent driver cells that do not trigger any operator; green circles represent driver cells that DO trigger an operator; grey circles represent left parts of never activated operators; brown circles represent right parts of operators, both active and inactive; rectangular boxes hold the CET values.}
\label{00Lintrees}
\end{center}
\end{figure}

%\colorbox{red}{L-systems}
Our model shares some similarities with L-systems. Both models have productions that replace existing symbols with other symbols: the key difference lies in the mechanism to generate new symbols. In an L-system the new symbols have to be listed explicitly, e.g.:
\begin{eqnarray}
a \rightarrow abbc, b \rightarrow cc
\end{eqnarray}
In our model the production specifies only the number of new symbols (proportional to the dimension of the development area), while the symbols themselves (the CET values) are generated through a fixed procedure (i.e. such procedure never changes and therefore does not need to be encoded in the genome). This characteristic is particularly important in an evolutionary perspective, because it allows a more compact representation of the productions in the genome, nevertheless able to generate many new symbols. Another important difference is that L-systems draw the symbols from a finite alphabet, while in our case the alphabet is virtually unbounded (an array of size 8 of 10-valued scalars has $10^8$ possible values). We believe that this ``unboundedness'' paves the way for open-ended evolution. 

%\colorbox{red}{comparison, Gruau}
Gruau's Cellular Encoding has a cell variable similar to the CET, but lacks the distinction between driver and normal cells. Due to this lack, each cell needs to be individually guided by the genome to develop, leaving the genetic algorithm with the impossible task to evolve as many operators as are the organism's cells (in the case of the human body, the genome would contain $10^{14}$ genes, instead of the $3*10^4$ it seems to have): for this reason, a CE-based system would not scale well to organisms with many cells. Moreover, in Gruau's model the cell state variable (the reading-head) is not a variable with an independent existence: it is a pointer on the tree of developmental instructions; such tree constitutes the main structure and a development operator cannot exist unless it coincides with a node of the tree. Our model on the contrary has a more flexible structure: there can be operators that are activated during development and operators that are never activated, CET values that activate an operator and CET values that never activate any operator. Both the CET values and the operators can move freely (by means of mutations) between the ``class of active elements'' and the ``class of inactive elements''.      

%\colorbox{red}{comparison, CA}
CA-based models of development also have a cell state variable and again the key difference lies in the mechanism of assignment of such value. While in CA-based models the value of the cell state is determined by the states (values of the same variable) of the surrounding cells, in our model it is assigned to cells when they are created, during a proliferation event. Of course this is not the only difference: in CA models there is no distinction between driver and normal cells either, etc. Actually, the main idea captured by CA-based models, which is intercellular communication, is not incompatible with our model. We can foresee an extension of it to include the influence of surrounding cells as determinants of cell behaviour, together with the CET value.  

%\colorbox{red}{comparison, cell chemistry}
In cell chemistry-based models of development the analog of the cell epigenetic type in our model is represented by the set of concentrations of certain chemicals which vary from cell to cell and trigger the activation of different genes in different cells: in this case the difference is again in the mechanism to generate new values. In cell chemistries no mechanism is present that guarantees that such ``memories'' are assigned different values in different cells; moreover, as the concentrations of chemicals can be influenced by many sources and are therefore quite ``volatile'' as memories, they are likely to change in the course of development, whereas the CET value is conceived as a non-volatile memory that, unless mutations occur, maintains its value during the entire development. A biologically plausible feature that many cell chemistry-based approaches possess is the presence of a genetic regulatory network, that allows cell behaviour to be determined by the interaction of many genes: also this idea is not incompatible with our model and it is in our plans to include it in future work.

\section{Experiments}

\begin{figure}[!t]
\begin{center}
\includegraphics[width=0.7\columnwidth]{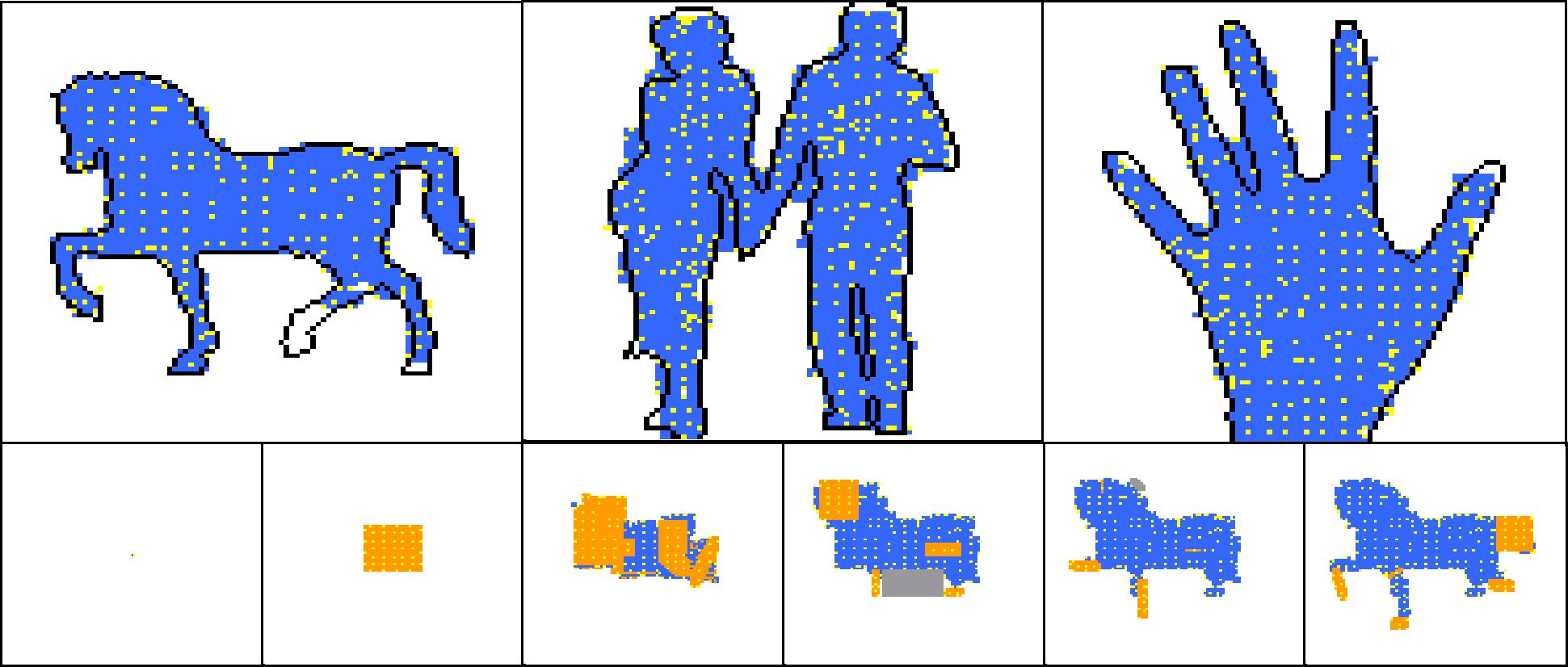}
\caption{The horse, the couple and the hand (dynamical view, best evolved shapes, target contour superimposed). In the lower part the development sequence of the horse.}
\label{00Frame00}
\end{center}
%\hskip 0.05cm
\end{figure}

\begin{table}[!t]
\center{
\begin{tabular}{|l|c|c|c|} \hline
statistics & horse & couple & hand \\ \hline\hline
image size & 100x100 & 100x100 & 100x100 \\
generations & 10720 & 3000 & 7280 \\
fitness value & 88.93 & 85.89 & 92.15 \\
dev. operators & 64 & 64 & 64 \\
operators used & 34 & 32 & 36 \\
CET created & 332 & 572 & 520 \\
genome size & 5635 & 5635 & 5635 \\ \hline
\end{tabular}}
\vskip 0.25cm
\end{table}

%\colorbox{yellow}{ga and niching}
The model of development described in the previous section has been tested on the problem of morphogenesis achieved by means of evolutionary techniques, i.e the task is to generate predefined 2-dimensional shapes by evolving genomes that guide the development of the shape starting from a single cell; the implicit assumption behind this choice is that a model of development is good if it is evolvable. The experimental procedure consists in evolving a population of genomes, at each generation letting the development unfold for each memory (starting from a single cell with CET = [0,...,0] placed in the middle of the grid and running GDP from 0 up to a maximum value), and then using the adherence of the shape at the end of development to the target shape as fitness measure. The genetic population is composed of 600 individuals (represented as strings of quaternary digits), undergoing elitism selection for up to 20000 generations. GA parameters are 50\% single point crossover, mutation rate of 0.1\% per digit. The fitness function formula is the same adopted by H. de Garis \citep{DG99}:
\begin{eqnarray}
F=(ins-outs)/des
\end{eqnarray}
where ins is the number of cells of the evolved shape falling inside the target shape, outs is the number of cells of the evolved shape falling outside the target shape, des is total the number of cells of the target shape. Other important parameters are (relevant reference values in parentheses):

\begin{figure}[!t]
\begin{center}
\includegraphics[width=0.7\columnwidth]{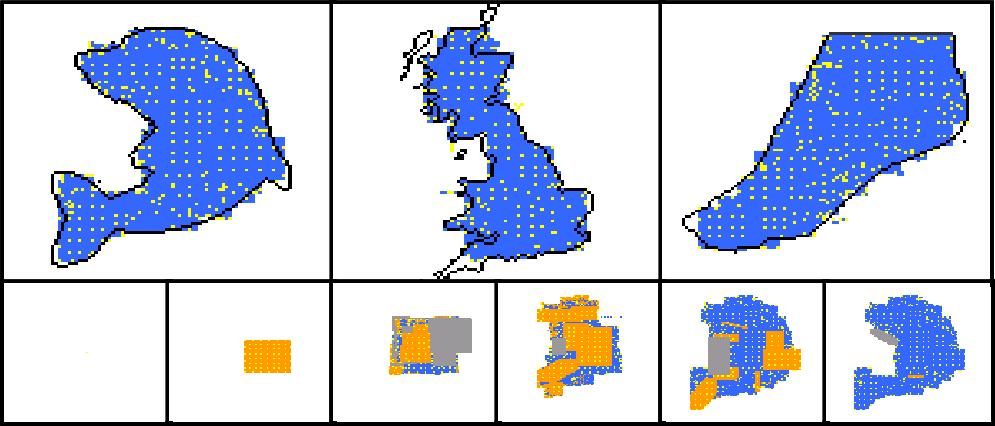}
\caption{The dolphin, the map of Britain and the foot (dynamical view, best evolved shapes, target contour superimposed). In the lower part the development sequence of the dolphin.}
\label{00Frame01}
\end{center}
\end{figure}

\begin{table}[!t]
\center{
\begin{tabular}{|l|c|c|c|} \hline
statistics & dolphin & Britain & foot \\ \hline\hline
image size & 100x100 & 100x100 & 100x100 \\
generations & 1580 & 1940 & 1560 \\
fitness value & 90.84 & 80.95 & 94.00 \\
dev. operators & 64 & 64 & 64 \\
operators used & 27 & 24 & 28 \\
CET created & 471 & 284 & 467 \\
genome size & 5635 & 5635 & 5635 \\ \hline
\end{tabular}}
\vskip 0.25cm
\end{table}

\begin{itemize}
\item number of development phases (8)
\item max no. of operators active in each phase (8)
\item normal to driver cells ratio in proliferation (25)
\item max linear dim. of development area (10-18 cells)
\end{itemize}

%\colorbox{yellow}{germline penetration}
As the number of possible epigenetic types (and hence the search space of the genetic algorithm) can be very large, the evolutionary process can become very slow. In order to overcome this problem, we introduced a mechanism called ``germline penetration'' by which (a subset of) the CET values that have occurred during the development of any given individual are converted into quaternary code and then copied into the XET variables of the operators in the genome as a ``suggestion'' for the genetic algorithm. The rationale behind this measure is the consideration that the only way to change the course of development of an individual is to introduce in the genome development operators acting on the epigenetic types that have occurred in (and hence have determined) the current development; otherwise most XET values in the left parts of the development operators would be wasted on values that never occurred. To avoid disrupting the current developmental path, the operators with the newly introduced epigenetic types are set as inactive.

%\colorbox{yellow}{views}
For each development phase two views are possible for the developing shape: one that shows the ``dynamics'' of the development operators, and one that shows the colours of cells. In the ``dynamical view'' driver cells are coloured in yellow, normal cells are coloured in orange if they have been just (i.e. in the current development phase) created, in blue if they have been created in one of the previous phases; areas where cells have been deleted by an apoptosis event are coloured in grey. In the ``colour view'' (which of course makes sense only for colour targets) cells are shown with their actual colours. In addition the contour of the target shape has sometimes been super-imposed. 

\begin{figure}[!t]
\begin{center}
\includegraphics[width=0.7\columnwidth]{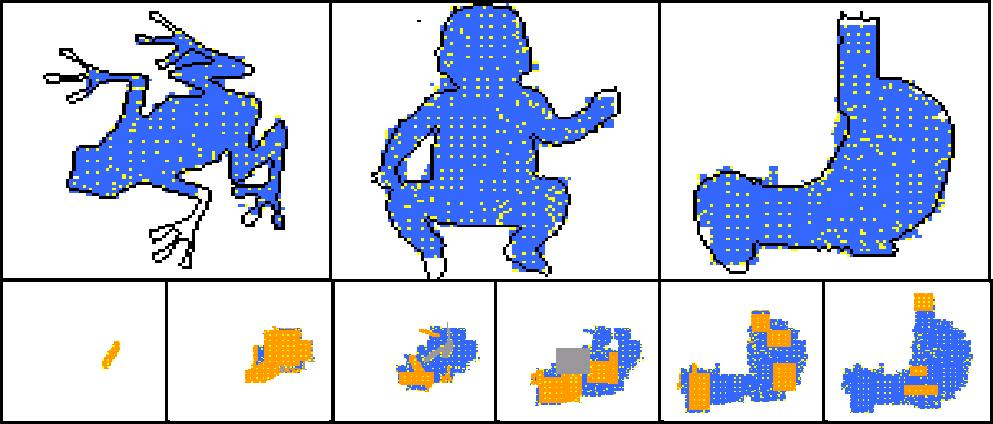}
\caption{The frog, the baby and the stomach (dynamical view, best evolved shapes, target contour superimposed). In the lower part the development sequence of the stomach.}
\label{00Frame02}
\end{center}
\end{figure}

\begin{table}[!t]
\center{
\begin{tabular}{|l|c|c|c|} \hline
statistics & frog & baby & stomach \\ \hline\hline
image size & 100x100 & 100x100 & 100x100 \\
generations & 5000 & 2320 & 1040 \\
fitness value & 79.01 & 84.44 & 90.06 \\
dev. operators & 64 & 64 & 64 \\
operators used & 26 & 27 & 25 \\
CET created & 420 & 444 & 420 \\
genome size & 5635 & 5635 & 5635 \\ \hline
\end{tabular}}
\vskip 0.25cm
\end{table}

%\colorbox{yellow}{targets}
Experiments have been conducted with a number of different target shapes; the targets have been chosen with the objective of testing the method on shapes as diverse as possible, to prove its effectiveness in generating any kind of shape. All targets are 100x100 multi-cellular arrays: the limited computational resources available prevented us from putting to a test larger shapes. Figures 7-11 show the results of experiments conducted with nine black-and-white targets (the horse, the couple, the hand, the dolphin, the map of Britain, the foot, the frog, the baby, the stomach) and two experiments conducted with colour targets (the french flag and the head).

%\colorbox{red}{general comment}
As we can see, all target shapes have been approximated to a good degree, with the exception perhaps of the frog. This is due to the fact that the frog's forelimbs are very thin (they contain a small number of cells), which raises the chances that a proliferation event creates cells falling outside the target shape, fact that is penalised by the fitness function; colour targets have proved more difficult to evolve, as one may have expected. To our knowledge, no other method is able to generate, by means of evolutionary techniques, target shapes of this dimension and this variety; the french flag presented in \citep{MB03}, for example, has a size of 40x40. Furthermore, the proposed method can be easily extended to the generation of 3D shapes. Another interesting observation concerns the percentage of inactive (hence unused) operators in the genome: it has been observed that if the percentage of such inactive operators falls below roughly 30\% of the total, evolution stops (i.e. fitness ceases to increase), and it only resumes when the dimension of the genome is increased, allowing for the presence of more inactive operators (e.g. bringing the number of operators in the genome from 48 to 64).

\section{Biological Interpretation}

%\colorbox{yellow}{biological grounds and morula}
The method presented can be interpreted as a model of embryogenesis. In this interpretation the biological counterpart of the epigenetic type could be implemented in real cells by means of either epigenetic or genetic mechanisms. In the first case it would be coded in the methylation patterns of chromosomes, in the second it would be coded in a portion of the genome reserved for this. In an article published recently \citep{TP07}, M. Zernicka-Goetz and some colleagues proved that the decision about the cell fate in the mouse embryo is taken as early as at the four-cell stage. This decision would be determined by the presence in the embryo cells of certain chemicals in asymmetrical quantities and would then be ``stored'' in the nucleus through changes in the histone-arginine methylation patterns of specific DNA segments. The information stored in such methylation patterns is the biological counterpart of the epigenetic type in our model. The first development steps always see the creation of a (usually rectangular) mass of cells: this mimics very closely the early development of the animal embryo, up to the morula stage. This similarity brings us to support the hypothesis that, even though they look the same, the morula cells already have different epigenetic types (i.e. the biological equivalent of CET), hypothesis that appears in accordance with \citep{TP07}.

\begin{figure}[!t]
\begin{center}
\includegraphics[width=0.60\columnwidth]{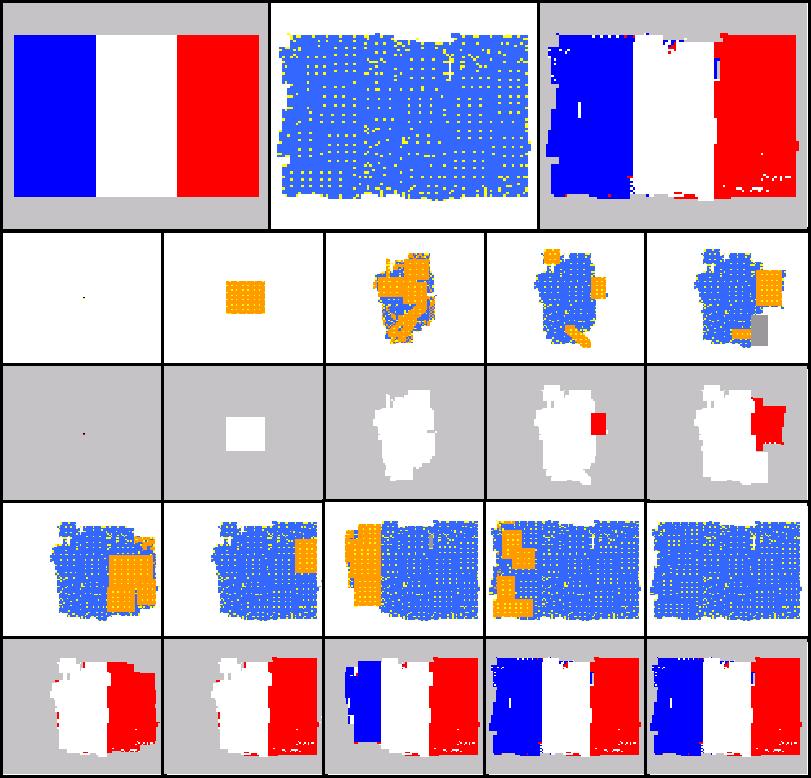}
\caption{The french flag. In the upper part, on the left the target shape, in the middle the best evolved shape in dynamical view and on the right the best evolved shape in colour view. In the lower part some development phases.}
\label{00French}
\end{center}
\end{figure}

%\colorbox{yellow}{stem cells, germline penetration, bio implausibilities}
The natural stem cells correspond to the driver cells: these are the cells that can receive guidance from the genome: all other cells are just steered by inter-cell signalling. The mechanism of germline penetration, which is very useful in reducing the search space of the GA, may well be present in nature and may be linked to the mobile DNA elements, whose presence is well documented since the pionerring studies of B. McClintock, and whose function is still largely unknown. At present we have no support for this hypothesis, but if this is true, it would correspond to a sort of  Lamarckian evolution, in that the developmental history of the organism influences the genome and therefore is passed on to the next generation. Speaking about things that lack biological plausibility, a feature of the proposed method which departs from what we know about biological systems is the fact the development events are triggered by operators that are activated by the CET of cells. In other words, we have a genetic regulatory network in which the outputs are directly connected to the input, i.e. a network with no ``hidden layer''. We know by contrast that in real cellular systems all events are determined by the interplay of many genes.

\begin{figure}[!t]
\begin{center}
\includegraphics[width=0.60\columnwidth]{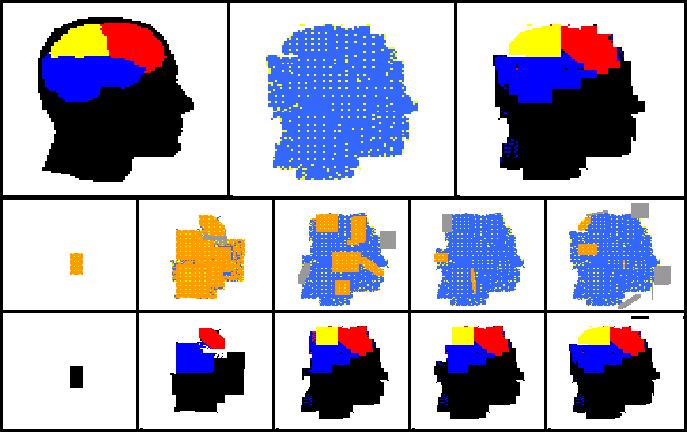}
\caption{The head. In the upper part, on the left the target shape and on the right the best evolved shape (colour view). In the lower part some development phases (dynamical view).}
\label{00Head}
\end{center}
\end{figure}

%\colorbox{yellow}{junk DNA}
In molecular biology, junk DNA is a collective label for the portions of the DNA sequence of a genome for which no function has yet been identified (non protein- or RNA-coding DNA); it includes most sequences within introns and most intergenic DNA. According to the model of development proposed, a development operator having an epigenetic type value that never shows up during the ``normal'' embryonic development, can be defined as ``junk'' genome. On the other hand all driver cells (i.e. their CET values) that are not activated by any operator during development can be as well termed ``junk'' driver cells (``junk'' CET values). By this definition, all CET values present at the end of development (the yellow dots in the snapshots showing the last development phase) are junk CET values. The presence of junk information in both the genetic and the epigenetic memory is consistent with the very nature of evolution, which is blind, devoid of any teleological aim and open-ended; it could be considered an unavoidable side-effect brought about by the use of evo-devo techniques.

%\colorbox{yellow}{junk DNA 2 and open-ended evolution}
If in a successive generation an existing junk operator is affected by a mutation that turns it into an operator that activates a junk CET value, both the junk operator and the junk CET value cease to be junk and give a contribution to the developmental process, that can continue and generate new lifeforms. On the other hand, if the embryonic development for whatever reason from departs normality and as a consequence new, unexpected CET values are created, development operators that would normally be inactive, can become active. These considerations suggest a more blurred scenario for the junk DNA, in which a given development operator can be junk or not depending on the actual conditions encountered by the organism during its development.

\section{Conclusion}

%\colorbox{yellow}{conclusion}
We presented a method to generate arbitrary shapes by using evolutionary-developmental techniques, that can also be interpreted as a model of embryogenesis and stem cells functioning. The method has been successfully tested on many black and white and colour target shapes, very different from each others, which brings us to say that the proposed method has general validity and is capable to generate any kind of shape. Future work includes modelling of inter-cellular signalling and replacing the current direct input-output connections with a multi-layered genetic regulatory network.

%\footnotesize

\end{document}